\documentclass[12pt]{article}
\usepackage{natbib}
\usepackage{amssymb}
\usepackage{epsfig}
\usepackage{aastex_hack}
\usepackage{deluxetable}
\usepackage{graphicx}
\date{}

\newcommand{\tildEta}{\eta}
\newcommand{\tildN}{\tilde{N}}
\newcommand{\tildP}{\tilde{P}}

\begin{document}

\title{Sub-Photospheric Emission from Relativistic Radiation Mediated Shocks in GRBs}
\author{Omer Bromberg$^1$, Ziv Mikolitzky$^2$ \& Amir Levinson$^2$\\
\small $^1$ Racah Institute of Physics, The Hebrew University, 91904 Jerusalem, Israel\\
\small $^2$ The Raymond and Berverly Sackler School of Physics and Astronomy,\\
\small Tel Aviv University, 69978 Tel Aviv, Israel}
\maketitle
%\opening

%
%
%%%%%%%%%%%%%%%%%%%%%%%%%%%%%%%%%%%%%%%%%%%%%%%%%%%%%%%%%%%%%%%%%%%%%
\begin{abstract}
%%%%%%%%%%%%%%%%%%%%%%%%%%%%%%%%%%%%%%%%%%%%%%%%%%%%%%%%%%%%%%%%%%%%%%
%
%

It is proposed that the prompt emission observed in bursts that exhibit a thermal component
originates from relativistic radiation mediated shocks that form below the photosphere of the GRB outflow.
It is argue that such shocks are expected to form in luminous bursts via collisions of shells that propagate
with moderate Lorentz factors $\Gamma\lesssim 500$.   Faster shells will collide above the photosphere to form collisionless shocks.
We demonstrate that in events like GRB 090902B a substantial fraction of the explosion energy is dissipated below the photosphere,
in a region of moderate optical depth $\tau\lesssim300$, whereas in
GRB 080916C the major fraction of the energy dissipates above the photosphere.
We show that under conditions anticipated in many GRBs, such relativistic radiation mediated shocks
convect enough radiation upstream to render photon production in the shock transition negligible,
unlike the case of shock breakout in supernovae.  The resulting spectrum, as measured in the shock frame,
has a relatively low thermal peak, followed by a broad, nonthermal component extending up to the KN limit.
\end{abstract}

%
%
%%%%%%%%%%%%%%%%%%%%%%%%%%%%%%%%%%%%%%%%%%%%%%%%%%%%%%%%%%%%%%%%%%%%%%
\section{Introduction}
%%%%%%%%%%%%%%%%%%%%%%%%%%%%%%%%%%%%%%%%%%%%%%%%%%%%%%%%%%%%%%%%%%%%%%
%
%

According to the standard view, the prompt GRB emission is produced behind internal shocks that form
in the coasting region of a baryon loaded fireball \citep[e.g.][]{LevEich93,ReesMes94,SariPiran97}.
The emission mechanism commonly invoked is synchrotron cooling of relativistic electrons
that are accelerated at the shock front, perhaps accompanied by synchrotron-self Compton emission at
highest energies. In order to be able to accelerate particles the shocks must be collisionless.  As
argued below (see also Levinson \& Bromberg 2008, hereafter LB08), under the conditions anticipated such
shocks can only form in regions where the optical depth for Thomson scattering is smaller than unity, as otherwise the radiation
produced in the shock transition and in the immediate downstream mediates the shock via Compton scattering.
Thus, an implicit assumption of the standard fireball model is that a significant fraction of the bulk energy dissipates
above the photosphere.

The fireball model outlined above was originally motivated by the detection of nonthermal spectra in many
GRBs.  Despite some difficulties in associating the low energy part of the spectrum with optically thin
synchrotron emission (e.g. Crider et al. 1997; Preece et al. 1998), some believed that this model
can account for the Band spectrum \citep{Band93} inferred in many sources.

The difficulties with the synchrotron model, as well as other considerations (e.g. Eichler \& Levinson 2000),
have led to the suggestion that photospheric emission may contribute to the prompt GRB spectrum, in addition to the
observed non-thermal component (e.g. M\'{e}sz\'{a}ros \& Rees 2000, M\'{e}sz\'{a}ros \textit{et al}. 2002).
At the same time, it has been argued that a combination of a thermal and a nonthermal components can fit better
the data in quite a number of bursts (e.g. Ryde 2004, 2005, 2006, Pe'er \textit{et al}. 2007, Pe'er 2008).
The recent detections of a prominent thermal peaks in several LAT bursts, notably GRB 090902B \citep{Abdo09b,Ryde10}, provide a
strong support to this interpretation.

The conditions at the shock formation region depend on the power and Lorentz factor of the outflow and the
duty cycle of the central engine.  For shells to collide above the photosphere, the
Lorentz factor must be large enough.  As shown below, in some bursts, e.g., GRB 080916C, internal shocks are expected
to form above the photospher, whereas in others, e.g., GRB 090902B, a major fraction of the energy is likely to dissipate
below the photosphere, albeit in a region of moderate optical depth.   Indeed, the spectrum of  GRB 080916C is well fitted by
a broken power law \citep{Abdo09a}, while that of GRB 090902B exhibits a thermal peak \citep{Abdo09b, Ryde10}.

The above considerations motivate detailed investigation of the properties of shocks that form below the photosphere,
in regions where the Thomson optical depth exceeds unity.  Preliminary studies indicate that such shocks
are mediated by Compton scattering of radiation produced in the shock transition layer or convected by the upstream flow
\citep[LB08;][]{Katz09, Budnik10}.  The typical width of the shock transition, a few
Thomson mean free paths, is much larger than any kinetic scale involved, rendering particle acceleration in such shocks
highly unlikely.  Thus, the photon spectrum produced in relativistic radiation mediated shocks (RRMS) may be considerably
different than that anticipated in
collisionless shocks.  Moreover, the radiation produced downstream is trapped for times much longer than the shock crossing time,
and diffuses out after the shock breaks out of the photosphere and becomes collisionless.  Consequently, any rapid
fluctuations of the flow parameters will be smeared out, and may not be imprinted in the observed gamma-ray emission.

In this paper we explore the role of RRMS in the prompt phase of GRBs.  A qualitative discussion of RMS is given in \S 2.
In \S 3 we analyze regimes in the parameter space
in which overtaking collisions of shells occur below and above the photosphere.   In \S 4 we investigate the properties
of RRMS that form during the prompt GRB phase.  We show that
under conditions anticipated in many GRBs, such relativistic radiation mediated shocks
convect enough radiation from the upstream to render photon production in the shock transition negligible,
unlike the case of shock breakout in supernovae and hypernovae (e.g., Weaver 1977; Katz et al. 2010).
Bulk Comptonization then produces a relatively low thermal peak, followed by a broad, nonthermal component
in the immediate downstream.   To illustrate some properties of the spectrum we present, in \S 5, test particle
Monte Carlo simulations of bulk Comptonization.
The shock structure computed in LB08 is used as input to these simulations.
In the regime where photon convection is important the analysis of LB08 is justified, and
the shock solutions obtained there present reasonable approximations.   We find that a hard tail extending
up to the KN limit (roughly  $\sim \Gamma m_e c^2$ in the observer frame) is a generic feature of RRMS in GRBs.
We conclude in \S 6.

%
%
%%%%%%%%%%%%%%%%%%%%%%%%%%%%%%%%%%%%%%%%%%%%%%%%%%%%%%%%%%%%%%%%%%%%%%
\section{RMS - a qualitative discussion and review of previous work}
%%%%%%%%%%%%%%%%%%%%%%%%%%%%%%%%%%%%%%%%%%%%%%%%%%%%%%%%%%%%%%%%%%%%%%
%
%

The structure of non-relativistic RMS was first studied by
\citet{Pai66}, \citet{ZR67} and by \citet{Colgate72}. These
authors assumed that the plasma and the radiation field maintain
local thermodynamic equilibrium (LTE) at the shock transition,
giving rise to a black body spectrum. Under this assumption the
temperature increases monotonically from the upstream to the
downstream. \citet{Weaver76} extended this analysis to include
conditions anticipated in supernovae shock breakout, where the
upstream is cold and photon poor. He showed that when the shock
velocity exceeds $\sim0.03$c and the upstream temperature
$KT_u\lesssim1~n_{b,15}^{1/3}$ eV,
where $n_{b,15}$ is the baryons mass in units of $10^{15}$ gr,
the photon production rate is insufficient to sustain a black body
distribution within the shock transition. As a consequence, the
photons can only maintain Wien equilibrium with the plasma,
defined as a Planck spectrum with a non-vanishing chemical
potential, and a temperature equal to that of the electrons. Since
fewer photons supply the shock pressure, the temperature rises
above its value at the far downstream, where thermodynamic
equilibrium is re-established. At $\beta_u\sim0.2$ the peak
temperature reaches values of $KT\sim10$ KeV, approximately two
orders of magnitude above the black-body temperature further
downstream. At such velocities the photon production rate in the
transition region is negligible, and the majority of photons which
support the shock are created downstream of the deceleration zone,
and diffuse back to the upstream \citep{Katz09}. Higher upstream
velocities lead to increased pair production- and annihilation
rates up to a point where the amount of MeV photons begins to
affect the flow. In this case, the full KN cross section must be
used to correctly solve the shock. RMS in this limit are analyzed
by \citep{Budnik10}.

When the photon-to-electron ratio in the upstream region is high enough, the
total production rate of photons
in the transition region (due to local emission processes and diffusion from the
downstream) is negligible;
%it may then be assumed that conservation of photon number prevails within the
it may be assumed that the photon number is conserved within the
shock. Blandford \& Payne (1981a,b; hereafter BP81a,b) investigated
such shocks in the limit of non relativistic upstreams and
vanishing thermal effects, and obtained a self-consistent solution
for the structure of the shock and for the transmitted radiation
spectrum. They showed that photons which are advected with the
flow across the shock, can diffuse back to the upstream and
traverse the shock once more. In each crossing the photons gain a
small amount of energy as a result of the flow compression. Some
of these photons are caught in the shock, and by experiencing this
process multiple times, undergo a substantial energy buildup,
resulting in a power law tail which extends to high energies,
similar to first order Fermi acceleration of cosmic rays
\citep[for a review see][]{BlandEich87}. The solution provided by
BP81b for the transmitted spectrum is valid as long as the average
energy gain per scattering due to compressional heating (bulk
motion), $\sim\beta^2$, sufficiently exceeds the corresponding
energy gain due to thermal Comptonization
$\sim\frac{KT_e}{m_ec^2}$. This condition is equivalent to the
requirement that the Compton parameter satisfies $y_C\lesssim1$,
and holds for fast, cold flows. BP81b solution was generalized by
\citet{LS82}, \citet{Riffert88} and \citet{Becker88} to include
thermal processes as well.

In the non-relativistic regime, where the upstream velocity $\beta_u\ll1$  and the
temperature across the shock $T \ll m_ec^2/K$,  several
approximations can be made which greatly reduce the complexity of
the solution. Firstly, the Thomson cross section $\sigma_T$ can be used in the
radiation transfer equation, instead of the full KN formula.
Secondly, as the radiation field anisotropy in the fluid rest
frame is always small, the motion of the photons across the shock
can be regarded as a diffusive process. In such process the number
of scatterings that a photon undergoes as it passes through the
shock deceleration region is $N=(L_s/\lambda_T)^2$, where $L_s$ is
the width of the deceleration region and $\lambda_T=(n_e\sigma_T)^{-1}$ is the photon mean
free path. If the diffusion time roughly equals to the shock crossing time then $\lambda_T
N\simeq L_s/\beta_u$. Consequently $N\simeq\beta_u^{-2}$ and
$L_s\simeq(n_e\sigma_T\beta_u)^{-1}$. Finally, to order
$\beta_{u}^2$, the radiation field satisfies the equation of state
$P_r = U_r/3$, which provides a closure condition for the set of
hydrodynamic equations governing the shock structure
\citep[e.g.][]{HS76}.

When $\beta_u$  approaches unity the photon distribution function becomes highly anisotropic and the diffusion
approximation breaks down.  A completely different method of solution is then required.   In addition, pair creation inside
the shock transition may be important  and needs a proper treatment.
LB08 computed the structure of the shock transition for various upstream condition, using multiple moments expansion
of the transfer equation.  They assumed that the photon number is roughly conserved across the shock transition,
which is justified when photon advection by the upstream fluid dominates over photon production, and neglected pair creation
by nonthermal photons.   They found, as expected,  that the thickness of the shock is of the order of a few Thomson lengths.
Recently, \citet{Budnik10} calculated numerically the structure and spectrum of RRMS for sufficiently relativistic, cold upstream.
They have shown that the spectrum consists of a thermal peak at $m_ec^2$ roughly, followed by a hard tail that extends
up to $\Gamma_u^2m_ec^2$, as measured in the shock frame.  The high temperature in the immediate downstream is a
consequence of the paucity of photons advected from the upstream,
resulting in intensive pair cascade at the immediate downstream which keeps the temperature at this level.
A much lower temperature is anticipated in the RRMS which are expected
to form in the photons rich jets of GRBs, as discussed below.

%
%
%%%%%%%%%%%%%%%%%%%%%%%%%%%%%%%%%%%%%%%%%%%%%%%%%%%%%%%%%%%%%%%%
\section{Formation of RRMS in GRB outflows}\label{implications}
%%%%%%%%%%%%%%%%%%%%%%%%%%%%%%%%%%%%%%%%%%%%%%%%%%%%%%%%%%%%%%%%
%
%

To investigate the conditions under which shells may collide below the photosphere
we consider a conical fireball having an isotropic equivalent
%To calculate the optical depth at the dissipation region, we
%consider a conical fireball having an isotropic equivalent
luminosity $L_{iso}$.   We assume that the fireball is  ejected
with an initial Lorentz factor $\Gamma_0\sim1$  from a compact central engine  of radius
$R_0$, and that it carries baryons with an isotropic mass loss rate $\dot M_b$.   The properties
of the fireball, and in particular the location of the photosphere depend of the dimensionless parameter
$\tildEta\equiv\frac{L_{iso}}{\dot M_bc^2}$.
When $\eta<\eta_c$, where
\begin{equation}
\tildEta_c=\left(\frac{\sigma_TL_{iso}\Gamma_0}{4\pi
R_0m_bc^3}\right)^{1/4}=
1.8\times10^3L_{52}^{1/4}R_6^{-1/4}\Gamma_0^{1/4},
\end{equation}
the fireball is sufficiently opaque, such that the radiation is trapped during the entire acceleration phase.
The major fraction of the explosion energy is
then converted into bulk kinetic energy of the baryons, and the fireball reaches a terminal Lorentz factor $\Gamma_\infty\simeq\eta$ at some
radius $r_*\simeq\eta R_0/\Gamma_0$, beyond which it continues to coast.
The photosphere is located somewhere in the coasting region, at $r_{ph}>r_*$.
On the other hand, when $\eta>\eta_c$  the fireball will become transparent already during the acceleration phase, before reaching the coasting
radius $r=\eta_c R_0/\Gamma_0$.   The Loerentz factor in that case is
limited by $\eta_c$
%by its value at the photosphere, roughly $(\eta_c/\eta)^{1/3}\eta_c$
\footnote {If $\tildEta>\tildEta_c$ the acceleration
continues beyond the photosphere, since the photon flux is high
enough to sustain efficient acceleration even though only a
fraction of the photons undergo scatterings \citep[see][]{Meszaros00, Nakar05}.}.

The optical depth at a radius $r$, defined as
$\tau(r)=\int_{r}^\infty\sigma_T n_l \Gamma^{-1}dr$,
can be expressed in terms of the fireball parameters as
\footnote{For simplicity we assumed an infinite medium. If the shock forms inside a finite
shell, the upper limit of the integration should change to $2r$
\citep[e.g.][]{ShemiPiran90, Meszaros93, Nakar05}, but that will
have only minor effect on our results.}
\begin{equation}\label{r_ph}
\tau(r)=\frac{\eta_c^4R_0}{\eta^3\Gamma_0 r} \left\{
\begin{array}{cc}
\left[3-2(\eta_c/\eta)^4\right]^{-1}(\eta R_0/\Gamma_0r)^2
&
\tildEta_c<\tildEta,\\
1&
\eta<\eta_c.
\end{array}
\right.
\end{equation}
The photospheric radius $r_{ph}$ can be found from the condition $\tau(r_{ph})=1$, and it is readily seen
that for $\eta=\eta_c$ one has  $r_{ph}=\eta_c R_0/\Gamma_0=r_*$.

Suppose now that intermittencies of the central engine lead to ejections of shells that collide at some
radius $r_d$.  Let  $\Gamma_1$ denote the Lorentz factor of a shell ejected at time $t_0$ and $\Gamma_2=b\Gamma_1$
the Lorentz factor of a second shell ejected at $t_0+\delta t$. If $b>1$ the shells
collide at $r_{d}\simeq2\Gamma_1^2c\frac{b^2}{b^2-1}\delta t$.  Now, if $\eta<\eta_c$ then the shells collide in the coasting region
and we have $\Gamma_1\simeq\eta$.  The optical depth at the radius of collision is obtained from Eq. (\ref{r_ph}):
\begin{equation}\label{tau}
\tau(r_d)=\frac{\eta_c^4}{2 \Gamma_0\tildEta^5}\left(\frac{c\delta t}{R_0}\right)^{-1}(1-b^{-2}),
\end{equation}
and it is seen that collision will occur below the photosphere, viz., $\tau(r_d)>1$ if
\begin{equation}\label{eta*}
\Gamma_1<\eta_{ph}\equiv360 L_{52}^{1/5}R_6^{1/5}(1-b^{-2})^{1/5}\left(\frac{c\delta t}{R_0}\right)^{-1/5}.
\end{equation}

As can be seen from Eq. (\ref{eta*}) it is easier for internal
shocks to form below the photosphere in bursts with high
luminosity. Such high luminous bursts with $L_{iso}\sim$ a few
$10^{53}$ ergs/s are detected now by the {\it Fermi} satellite.
These bursts are characterized by an initial prompt phase
with a maximal energy $\lesssim10$ MeV and a peak energy at about 1
MeV, followed by a longer phase characterized by a harder spectrum
extending up to 10 GeV and a larger peak energy. The Lorentz
factors, inferred from the requirement that the shell should be
optically thin for the GeV photons, are usually high with
$\Gamma>500$, but these are model dependent
\citep[see e.g.][for lower values]{Zou10}.
%(see e.g. Zou et al. 2011 for lower values).
Table 1 shows some relevant values for 3
such bursts, where $\Gamma_{min,1}$ is the minimal estimated
Lorentz factor assuming two separate emission zones for the MeV
and for the GeV photons \citep{Zou10}, and $\Gamma_{min,2}$ is the
minimal Lorentz factor assuming a single emission zone.

\begin{figure}[h]
\begin{center}$
\begin{array}{c}
\begin{array}{cc}
\hspace{-10mm}
\includegraphics[width=3in]{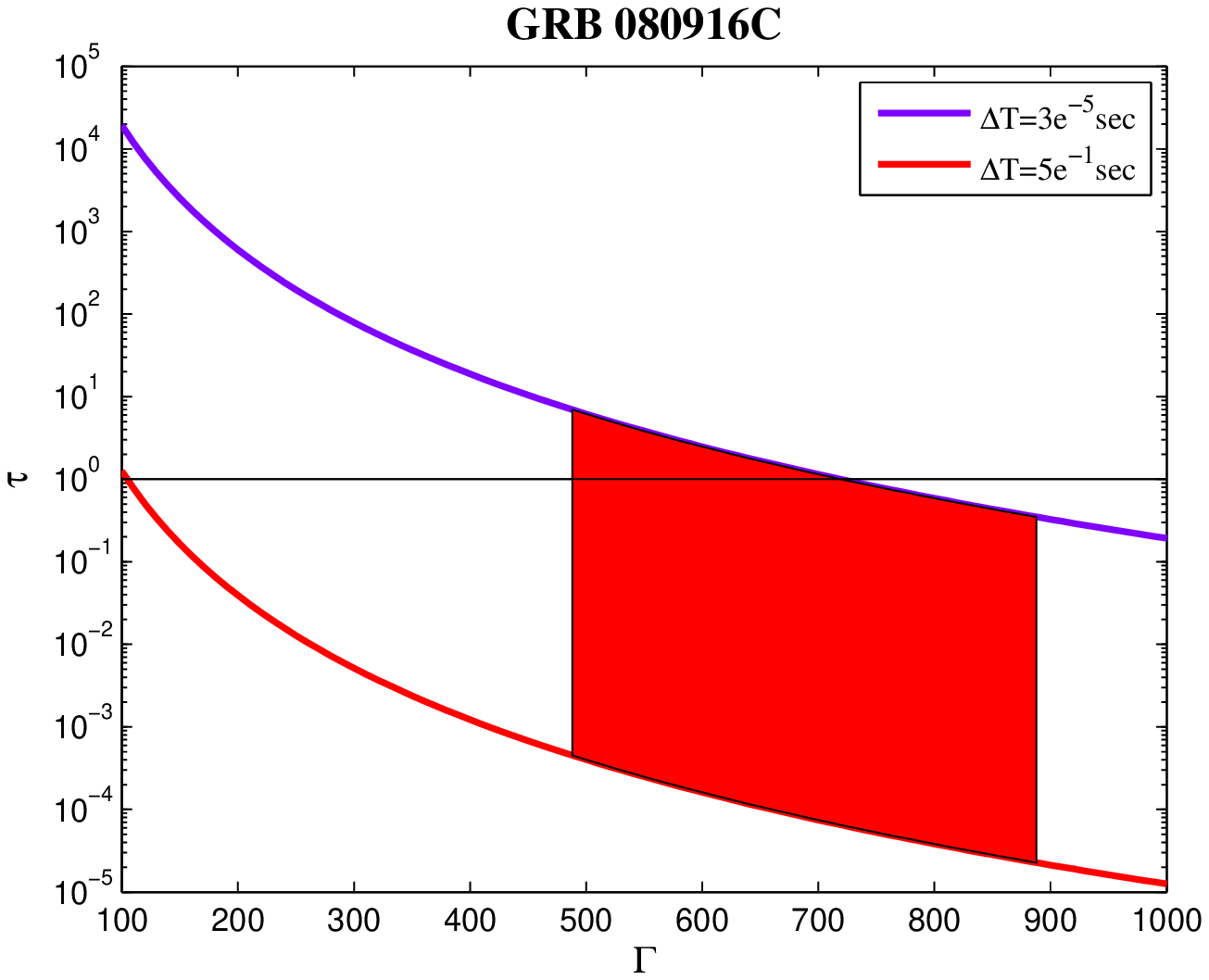} &
\hspace{-10mm}
\includegraphics[width=3in]{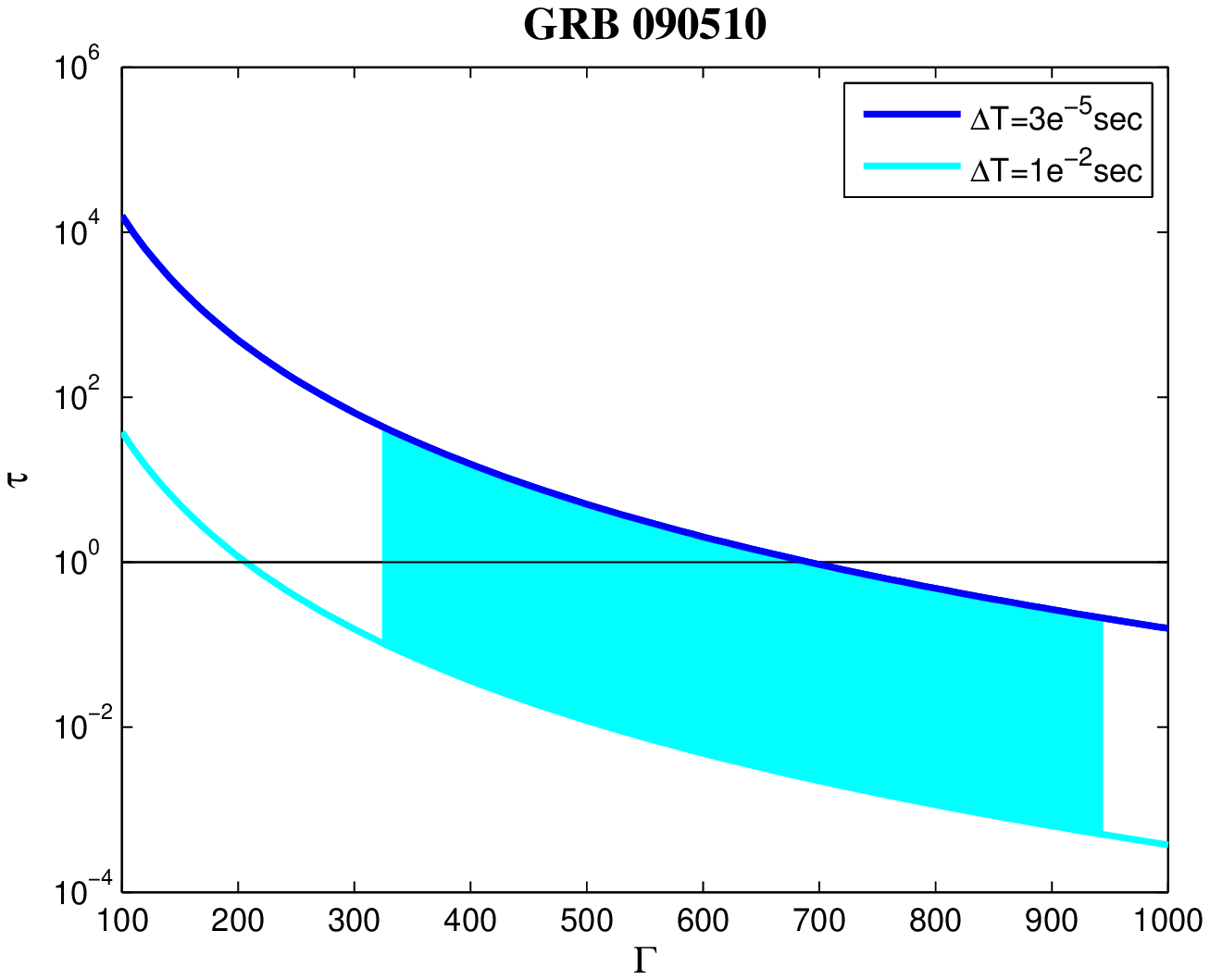}
\end{array}
\\
\hspace{-10mm}
\includegraphics[width=3in]{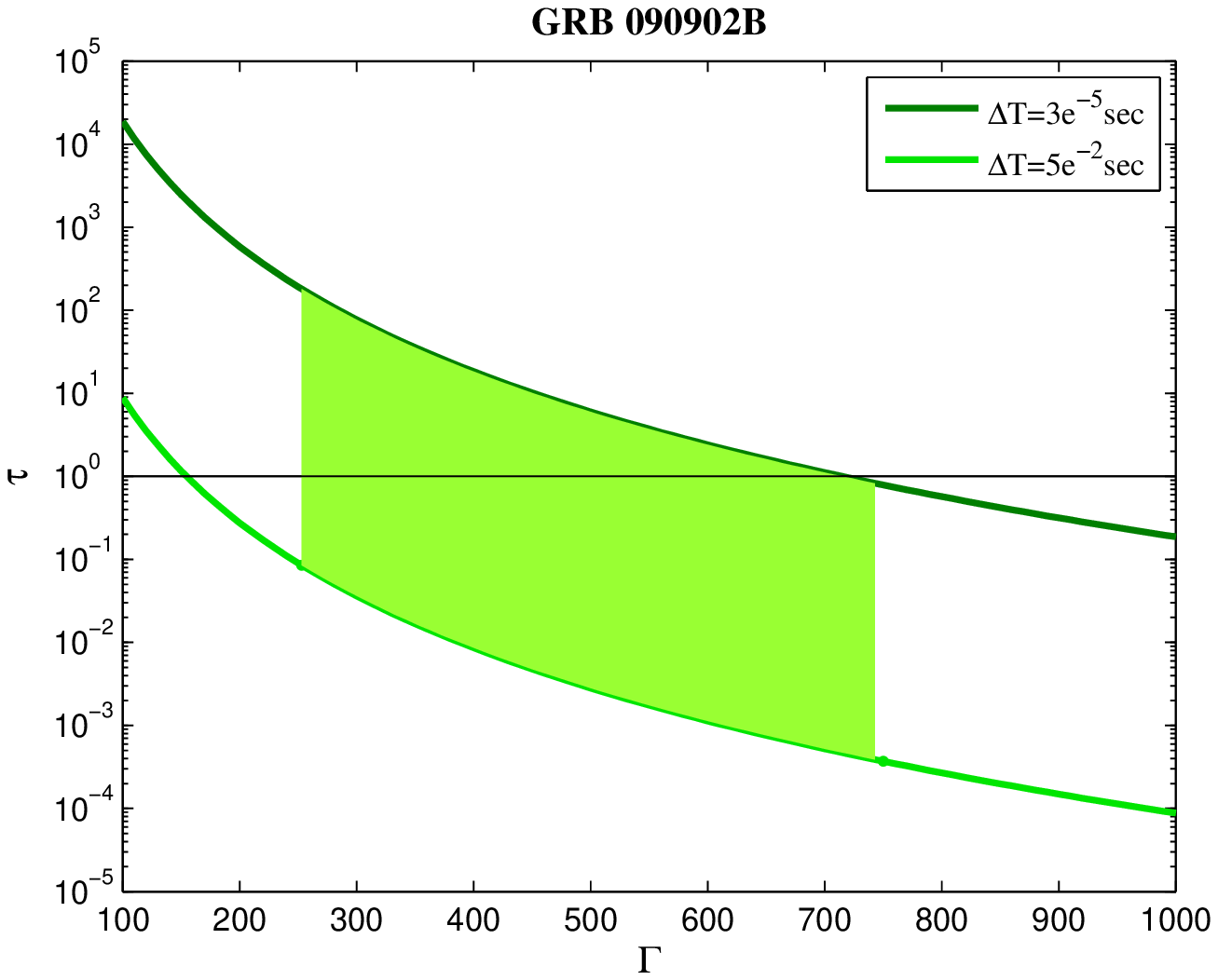}
\end{array}$
\end{center}
\caption{The optical depth at the shock location calculated for
different Lorentz factors and time separation of the ejected
shells, in 3 Fermi GRBs. The curve lines depict constant $\delta
t$, taken between a minimal dynamic time scale to a maximal value
taken to be the observed variability by {\it Fermi} BGO detector.
The colored areas show the range of $\Gamma$ and $\delta t$
relevant for each burst.}\label{Fermi_GRB}
\end{figure}

In fig (\ref{Fermi_GRB}) we plot $\tau(r_d)$ as a function of the
shell's Lorentz factor, for the 3 bursts shown in table 1. The
bold lines depict constant $\delta t$, and span a range between
$R_0/c$ and a maximal time interval, taken to be the observed
variability time by the {\it Fermi} BGO detector. The
colored
%shaded
areas mark the range of $\Gamma$ and $\delta t$ relevant for each
burst. As can be seen shells with $\Gamma>700$ collide above the
photosphere, resulting in shocks which are most likely
collisionless. On the other hand, slower shells ejected over time
scales which are closer to the dynamical time scale of the system,
collide below the photosphere leading to the formation of RMS. The
parameter space explored in fig (\ref{Fermi_GRB}) indicates that
in GRB 080916C, the majority of the GRB energy dissipates above
the photosphere, therefore a thermal component may be very week or
absent. In GRB 090902B, however a major fraction of the energy is
likely to dissipate below the photosphere, in a region of moderate
optical depth ($\tau<300$). Indeed a prominent thermal component
in the prompt soft phase accompanied by a hard tail was reported
in this burst \citep{Abdo09b}, as expected in a case of RRMS (see
\S\ref{RRMS.spectrum}). In the short GRB 090510, a moderate
fraction of the jet energy may dissipate below the photosphere.
The interpretation of the soft initial phase in this burst however
is not clear and may fit a Band spectrum such as expected from
Collisionless shocks, as well as a photospheric emission with a
(quasi) thermal component \citep[see e.g.][]{Ackermann10,
Pelassa10}.

Note that much longer variability time scale is
anticipated in a case where the shock forms below the photosphere,
and the downstream radiation is trapped for times much longer than
the shock crossing time, and diffuses out after the shock breaks
out of the photosphere. Therefore the observed variability timescales do not
necessarily reflect the actual time separation
between the shells, which may be much shorter. At what depth
downstream of the shock equilibrium is established is yet an open
issue. Since the enthalpy downstream is dominated by radiation, a
full equilibrium is not expected for the moderate optical depths
found above (see further discussion in \S\ref{RRMS.spectrum}). The
photons produced by RMS can reach energy of $\gtrsim\Gamma
m_ec^2=0.25\left(\frac{\Gamma}{500}\right)$ GeV in the observer
frame, which is more then enough to account for the energy range
observed during the soft phase.

\begin{table}[ht]
\caption{Parameters of Fermi LAT bursts}
\begin{center}
% use packages: array
\begin{tabular}{lllllll}
\hline GRB & $T_{90}$(s) & $E_{iso}$(erg) & $L_{iso}$(erg/s) &
$\delta t_{var}^*$ (ms))
& $\Gamma_{min,1}$ & $\Gamma_{min,2}$ \\
\hline $080916C^1$ & 66 & $8.8\cdot10^{54}$ & $3.3\cdot10^{53}$ &
$512$ & $488^2$ &
$890^1$ \\
$090510^3$ & 0.5 & $1.1\cdot10^{53}$ & $2.7\cdot10^{53}$ & $14$ &
$324^2$ &
$720^3$ \\
$090902B^4$ & 30 & $3.63\cdot10^{54}$ & $2.4\cdot10^{53}$ & $53$ &
$253^2$ &
$550^5$ \\
\hline
\end{tabular}
\end{center}
\small{ (1) \citet{Abdo09a}; (2) \citet{Zou10}; (3)
\citet{Ackermann10}; (4) \citet{Abdo09b}; (5)\citet{Ryde10} }
\newline
* The variability is measured with the BGO detector (150 Kev - 40 MeV).

\end{table}

%
%
%%%%%%%%%%%%%%%%%%%%%%%%%%%%%%%%%%%%%%%%%%%%%%%%%%%%%%%%%%%%%%%%%%%%%%%%
\section{Properties of RMS with photon rich upstreams}
%%%%%%%%%%%%%%%%%%%%%%%%%%%%%%%%%%%%%%%%%%%%%%%%%%%%%%%%%%%%%%%%%%%%%%%%
%
%

Consider an infinite plane parallel RMS moving in the $x$ direction.
The shock transition is defined within
a region bounded by $x_s-L_s$ and $x_s$, where $\beta=\beta_u$ if
$x<x_s-L_s$ and $\beta=\beta_d$ if $x>x_s$.
%{\bf The subindex $u(d)$ represent quantities at the upstream (downstream) of the shock}.
Far from the transition region the photons maintain thermodynamic equilibrium
with the plasma, and have a black body distribution with an
average energy per photon $\left<h\nu'\right>_{u(d)}=3KT'_{u(d)}$.
Hereafter primed quantities are measured in the fluid rest frame
and unless stated otherwise, unprimed quantities refer to the
shock frame.
We define the shock upstream to be photon-rich,
if the ``effective'' increase in the photon number
when passing through the shock transition is small.
The term "effective" is used to indicate that only those
photons which significantly contribute to the pressure are
accounted for \citep[see similar analysis in][but in a different regime of parameters spance]{Katz09}.
Taking $T'_s\equiv T'(x_s)$ as the typical temperature
at the shock transition\footnote {In shocks where bulk
acceleration is significant, the spectrum is non-thermal; defined
as such, this temperature represents the ratio between the
radiation pressure and the photon number density. Note that the
peak temperature in the shock may be higher than $T'_s$, yet we
find that in the parameter regime relevant here, the difference
never exceeds an order of a few.}, photons which are produced with
energies $\ll KT'_s$ can also contribute to the radiation
pressure, if they can be IC upscattered to energies
of $\sim KT'_s$ during the passage through the shock. In
non-relativistic RMS, where thermal Comptonization is the dominant
heating mechanism, a photon can double its energy after
$\sim\frac{m_ec^2}{4KT'_s}$ scatterings. Therefore, the lowest
frequency above which the photon energy can significantly increase
is set by the condition
$\alpha_{ff}\lambda_T\frac{mc^2}{4KT'_s}<1$, where $\alpha_{ff}$
is the free-free absorption coefficient \citep[e.g.][]{Rybicki79},
giving
$h\nu'_{min}\sim50~eV~n'^{0.5}_{15}\left(\frac{KT'_s}{100eV}\right)^{-5/4}$.
Such a photon can reach the peak energy within the shock crossing
time if $y_C>ln\left(\frac{KT'_s}{h\nu'_{min}}\right)$
\citep{Weaver76, Katz09}. If,
however, bulk acceleration is the dominant mechanism,
$y_C\simeq1$, and only photons with energies above
$h\nu'_{min}\sim KT'_s/2$ can contribute to the pressure. When
$\gamma_u\gg1$ each photon undergoes only a few scatterings,
experiencing an energy increase $\lesssim\gamma_u^2$. Therefore
$h\nu'_{min}\sim KT'_s/\gamma_u^2$ at this limit.

In addition to the Comptonization processes, emission can also be an
important source of photons if the
photons production time is shorter than the shock crossing time.
\citet{Svensson84} demonstrated
that the two most dominant emission processes in a thermal plasma,
when pair density is negligible, are e-p Bremsstrahlung and double
Compton (DC). The effective photon generation rate by Maxwellian
e-p Bremsstrahlung emission is:
\begin{equation}\label{RMS.ff}
\dot{n}'_{ff}=\sqrt{\frac{32\pi}{3}}\left(\frac{T'}{m_ec^2}\right)^{-1/2}(n'_l\sigma_Tc)n'_b\alpha\Lambda_{ff},
\end{equation}
where $\Lambda_{ff}\equiv\left(-{\rm
Ei}(\frac{h\nu'_{min}}{m_ec^2})\right)\cdot g_{ff}$, ${\rm Ei}(x)$
is the exponential integral of $x$, and $g_{ff}$ is the Gaunt
factor \citep[e.g.][]{Rybicki79}. For the above estimations of
$h\nu'_{min}$ it is safe to assume that $\Lambda_{ff}\lesssim10$
for most upstream conditions. Here and for the rest of the paper,
electrons and protons are assumed to have a Maxwellian
distribution. This assumption relies on the fact that the
relaxation time of an electron in restoring a Maxwellian
distribution with temperature $T_e$ is of the order of
$t_e\sim(n_l\sigma_Tc~\rm{ln}\lambda)^{-1}(kT_e/m_ec^2)^{3/2}$,
where $\lambda$ is the Coulomb logarithm
\citep[see][BP81a]{Spitzer78}. For electrons with subrelativistic
temperatures this time is much shorter than the average time
between scatterings, $t_c\sim(n_l\sigma_Tc)^{-1}$.
\citet{Svensson84} calculated the production rate of photons by DC
emission in the limit of soft photons
$\left(h\nu'\ll\min(m_ec^2,KT')\right)$, having a Wien
distribution and maintaining thermal equilibrium with electrons.
He showed this rate to be
\begin{equation}\label{RMS.DC}
\dot{n}'_{DC}=\frac{4}{3\pi}n'_l\sigma_Tc\alpha\frac{\left<h^2\nu'^2\right>}{m_ec^2}n'_{r}\Lambda_{DC},
\end{equation}
where $n'_r$ is the photon number density,
$\Lambda_{DC}=\ln\left(\frac{KT'_s}{h\nu_{min}}\right)\bar{g}_{DC}$, and
$\bar{g}_{DC}$ is a numerical factor which replaces the gaunt factor.
We stress that the photons within the transition region may not be in
local equilibrium with the electrons \citep[e.g.][]{BP81b, Riffert88};
nevertheless, \citet{Chluba07} showed that if the average photon energy
exceeds that of the electrons, the emission efficiency is reduced.
Consequently, eq. (\ref{RMS.DC}) can be used as an upper limit.

Last, diffusion from the downstream provides an additional source
of photons to the shock transition region. Downstream of $x_s$ the
photon number density continues to rise due to emission processes,
until the photons regain thermodynamic equilibrium with the
plasma. This leads to propagation of photons backward toward the
upstream in a manner which can be treated as a diffusive process
as long as
%$\frac{\dot{n}'_r}{n'_r}<<(\sigma_Tn'_{l,d}\beta_dc)$.
${\dot{n}'_r}/{n'_r}<<(\sigma_Tn'_{l,d}\beta_dc)$.
The distance, $L_D$, which photons can cover by diffusion against
the flow before being swept back, can be estimated by maximizing
the expression $L_D= l_D-\beta_dct$, where $l_D\equiv2\sqrt{Dt}$
is the diffusion length and $D=\frac{c}{3n'_l\sigma_T}$ is the
diffusion coefficient \citep[e.g.][BP81a]{Weaver76}. This gives
$L_D=\frac{1}{3n'_{l,d}\sigma_T\beta_d}$, implying that photons
can diffuse back as long as $\beta_d<1/3$, which always holds in a
shock downstream. The total number density of photons at point
$x_s$ can thus be estimated by summing the total number of photons
swept from the upstream with those which diffuse back from the
downstream.
\begin{equation}\label{RMS.n_ph}
n'_{r,s}=\frac{n'_{r,u}u_u}{u_d}\left(1 + \frac{\int_{x_s-L_s}^{x_s+L_D}\dot{n}'_rdx}{n'_{r,u}u_u}\right)\equiv
\frac{n'_{r,u}u_u}{u_d}\left(1+\zeta\right),
\end{equation}
where $\dot{n}'_r\equiv\dot{n}'_{ff}+\dot{n}'_{DC}$.
The dimensionless parameter $\zeta$ quantifies the relative contribution of photon production inside the shock
and at the immediate downstream.
Specifically, $\zeta\ll1$ implies that the net production rate is much smaller than the incoming flux of photons,
so that the radiation field in the shock transition is dominated by
photons advected from the upstream.
The upstream in this case is defined to be photon-rich.
Relativistic temperatures or the presence of sufficient amount of
pairs introduce additional emission processes such as  pair
creation and annihilation and $e_+e_-$ Bremsstrahlung. However,
as shown below, our analysis is restricted to situations
where pairs are unimportant and temperatures are well below
$m_ec^2$, therefore this possibility is ignored.

An upper limit on $\zeta$ can be derived analytically using the following assumptions:
\begin{enumerate}
\item Within the shock transition the photon flux is conserved to a good approximation, i.e.
$n'_{r,u}u_u=n'_{r,d}u_d$ and
\begin{equation}\label{RMS.Tmax}
 KT'_s=P'_{r,d}u_d/n'_{r,u}u_u.
\end{equation}
\item The photon production rate at the shock transition and the immediate downstream is constant and equals
to the peak production rate of the shock at $x_s$.
\item The photons at $x_s$ have a Wien distribution with a temperature that is equal to that of the electrons. In this limit
$\left<h^2\nu'^2_{s}\right>=12\left(KT'_s\right)^2$.
\end{enumerate}
The upper limit of $\zeta$ resulting from these assumptions is:
\begin{equation}\label{RMS.zeta1}
\zeta_c=\frac{n'_{l,d}\alpha}{n'_{l,u}\gamma_u\beta_u^2}
\left(
0.02\lambda_c^3n'_{b,d}\Theta'^{-3}_u\Theta'^{-0.5}_s\Lambda_{ff}+
7\Theta'^{2}_s\frac{n_{r,d}}{n_{r,u}}\Lambda_{DC}
\right),
\end{equation}
where $\Theta'\equiv\frac{KT'}{m_ec^2}$. As it is shown below each
term in the brackets becomes important at different regimes of
upstream conditions. The first term is generally more important in
the non relativistic limit, while the second term becomes
important when the upstream has relativistic velocities.

In the limit where emission processes are negligible, the photon spectrum in the
shock transition and in the immediate downstream is determined by
a combination of thermal and bulk Comptonization.
When thermal Comptonization is the dominant process, the photons establish
a kinetic equilibrium with the electrons, resulting in a Wien spectrum.
A hard non-thermal tail may evolve only when photons are
upscattered by the bulk motion of the fluid at the upstream, and cross the shock
before being thermalized.
For a population of electrons with a non-relativistic temperature, the
photon thermalization time is  $t_T\sim\left(n'_b\sigma_Tc\frac{4KT'_s}{m_ec^2}\right)^{-1}$.
The shock crossing time is estimated differently for non-relativistic and
relativistic upstreams, where in the non-relativistic case
$t_s\sim\left(3n'_{b}\sigma_Tc/\beta_u^{2}\right)^{-1}$. If however
the upstream is relativistic
$t_s\sim\left(n'_{b}\sigma_Tc\tau/\gamma_u^2\right)^{-1}$,
where $\tau$ is the optical depth of the shock transition for a photon
coming from the {\it downstream}.
LB08 calculated $\tau$ to be of the order of a few ($\tau\sim3-5$),
implying that a mutual parameter for both cases can be defined,
$\Upsilon\equiv\frac{t_s}{t_T}\sim12\frac{KT_s'}{m_ec^2}\frac{1}{\gamma_u^2\beta_u^2}$.
If $\Upsilon<1$,
%implying
%that when the parameter $\Upsilon\equiv\frac{t_s}{t_T}\sim12\frac{KT_s'}{m_ec^2}\frac{1}{\gamma_u^2\beta_u^2}<1$,
a photon
that is upscattered at the upstream can cross the shock without loosing its energy
to thermal scattering, and a high energy tail may evolve.
Numerical simulations we preformed show that in the limit of $\Upsilon\rightarrow 0$,
this tail is hard and reaches energies up to $\gamma_um_ec^2$ in the shock frame
(see \S\ref{RRMS.spectrum}).
%As we show in \S\ref{RRMS.spectrum}, in the limit of $\Upsilon\rightarrow 0$, this tail
%reach energies up to $\gamma_um_ec^2$ in the shock frame.

%%%%%%%%%%%%%%%%%%%%%%%%%%%%%%%%%%%%%%%%%%%%%%%%%%%%%%%%%%%%%%%%%%%%
\subsection{Limits on the upstream conditions}\label{RRMS.conserveNR}

In order to analyze the conditions under which
the shock upstream can be considered photon-rich,
we express the downstream quantities in
eq.(\ref{RMS.zeta1}) in terms of the upstream values. Those are
derived from the fluid equations, assuming conservation of total
mass, energy and momentum across the shock. When $\beta_u\ll1$ the
shock jump conditions can be written as:
\begin{eqnarray}
n'_{b,u}\beta_u=n'_{b,d}\beta_d,\label{eq.beta.NR}\\
\left(n'_{b,u}m_bc^2\frac{\beta_u^2}{2}+\frac{4}{3}P'_{r,u}\right)\beta_u=
\left(n'_{b,d}m_pc^2\frac{\beta_d^2}{2}+\frac{4}{3}P'_{r,d}\right)\beta_d,\label{eq.Energy.NR}\\
n'_{b,u}m_pc^2\beta_u^2+P'_{r,u}=n'_{b,d}m_pc^2\beta_d^2+P'_{r,d}\label{eq.P.NR},
\end{eqnarray}
Here we assume that at the upstream and downstream the
contribution of baryon pressure is negligible. Such assumption is
easily justified in photon-rich upstreams,
since the number density of photons far
exceeds that of the electrons and protons. In terms of the
upstream parameters we therefore get:
\begin{eqnarray}
\beta_d=\frac{\beta_u}{7}\left(1+8\frac{P'_{r,u}}{n'_{b,u}m_pc^2\beta_u^2}\right),\\
P'_{r,d}=\frac{6}{7}n'_{b,u}m_pc^2\beta_u^2\left(1-\frac{1}{6}\frac{P'_{r,u}}
{n'_{b,u}m_pc^2\beta_u^2}\right)\label{RMS.jumpNR.P}.
\end{eqnarray}
Note that the radiation pressure in the upstream is limited by
$P'_{r,u}<\frac{3}{4}n'_{b,u}m_pc^2\beta_u^2$. Above this limit
$\beta_u$ is below the soundspeed, and no shock can form. We now
substitute $P'_{r,u}$ and $n'_{b,u}$ with the dimensionless
parameters $\tilde{P}\equiv\frac{P'_{r,u}}{n'_{b,u}m_pc^2}$ and
$\tilde{N}\equiv\frac{n'_{r,u}}{n'_{b,u}}$, which relate to the
original upstream parameters through:
\begin{eqnarray}
KT'_u&\simeq&90~eV~\tilde{P}_{-3}\tilde{N}_4^{-1},\\
n'_{b,u}&\simeq&2.4\times10^{15}~cm^{-3}~\tilde{P}_{-3}^3\tilde{N}_4^{-4}.
\end{eqnarray}
The resulting shock characteristic temperature in the
non-relativistic limit is:
\begin{equation}\label{Ts_NR}
KT'_{s,NR}=0.2~KeV~\tilde{N}_{4}^{-1}\beta_{-1}^2\frac{1+0.8\tilde{P}_{-3}
\beta_ { -1 }^{-2}}{1.8}
\end{equation}
and consequently, the derived upper limit for the relative number of newly
created photons within the shock is:
\begin{equation}\label{Zeta_NR}
%\small{
\zeta_{NR}=0.7\tilde{N}_4^{-0.5}\beta_{-1}^{-3}\Lambda_{ff,1}
%\frac{\Lambda_{ff,1}}{\tilde{N}_4^{0.5}\beta_{-1}^{3}}
\left(\frac{1+0.8\tilde{P}_{-3}\beta_{-1}^{-2}}{1.8}\right)^{-2.5}
+
1.2\times10^{-4}\tildN_4^{-2}\beta_{-1}^2\Lambda_{DC,1}.
%\left[1+1.7\cdot10^{-4}\frac{\beta_{-1}}{\tilde{N}_4^{1.5}}\frac{\Lambda_{DC,1}}{\Lambda_{ff,1}}\left(\frac{1+0.8\tilde{P}_{-3}\beta_{-1}^{-2}}{1.8}\right)^{1/2}\right].
%}
\end{equation}
Since both parameters are defined in terms of the upstream values we dropped the subscript $u$ for convenience.

When $\gamma_u\beta_u>1$, $T_s'$ and
$\zeta_c$ can be estimated by solving the relativistic generalization of
eqn. (\ref{eq.beta.NR}-\ref{eq.P.NR})
and taking  $\beta_d\sim1/3$.
This gives:
%\begin{equation}
%P'_{r,d}=\frac{2}{3}n'_{b,u}m_pc^2\gamma_u\left(1+4\tilde P\right)\label{RMS.jumpR.P},
%\end{equation}
%which results in a shock temperature
%
\begin{equation}\label{Ts_R}
KT'_{s,R}\simeq23~KeV~\tilde{N}_5^{-1}\gamma_{1}\left(1+0.04\tilde{P}_{-2}\right),
\end{equation}
and
\begin{equation}\label{Zeta_R}
\zeta_R\simeq0.1\tilde{N}_5^{-2}\gamma_{1}^3\Lambda_{DC,1}\left(1+0.04\tilde{P}_
{ -2}\right)^{-2}
\left[
1+0.04\frac{\tilde{N}_5^{1.5}}{\gamma_{1}^{2.5}}\frac{\Lambda_{ff,1}}{\Lambda_{
DC ,1}}\left(1+0.04\tilde{P}_{-2}\right)^{1.5}
\right].
\end{equation}

\begin{figure}[h]
\begin{center}$
\begin{array}{cc}
\hspace{-10mm}
\includegraphics[width=3in]{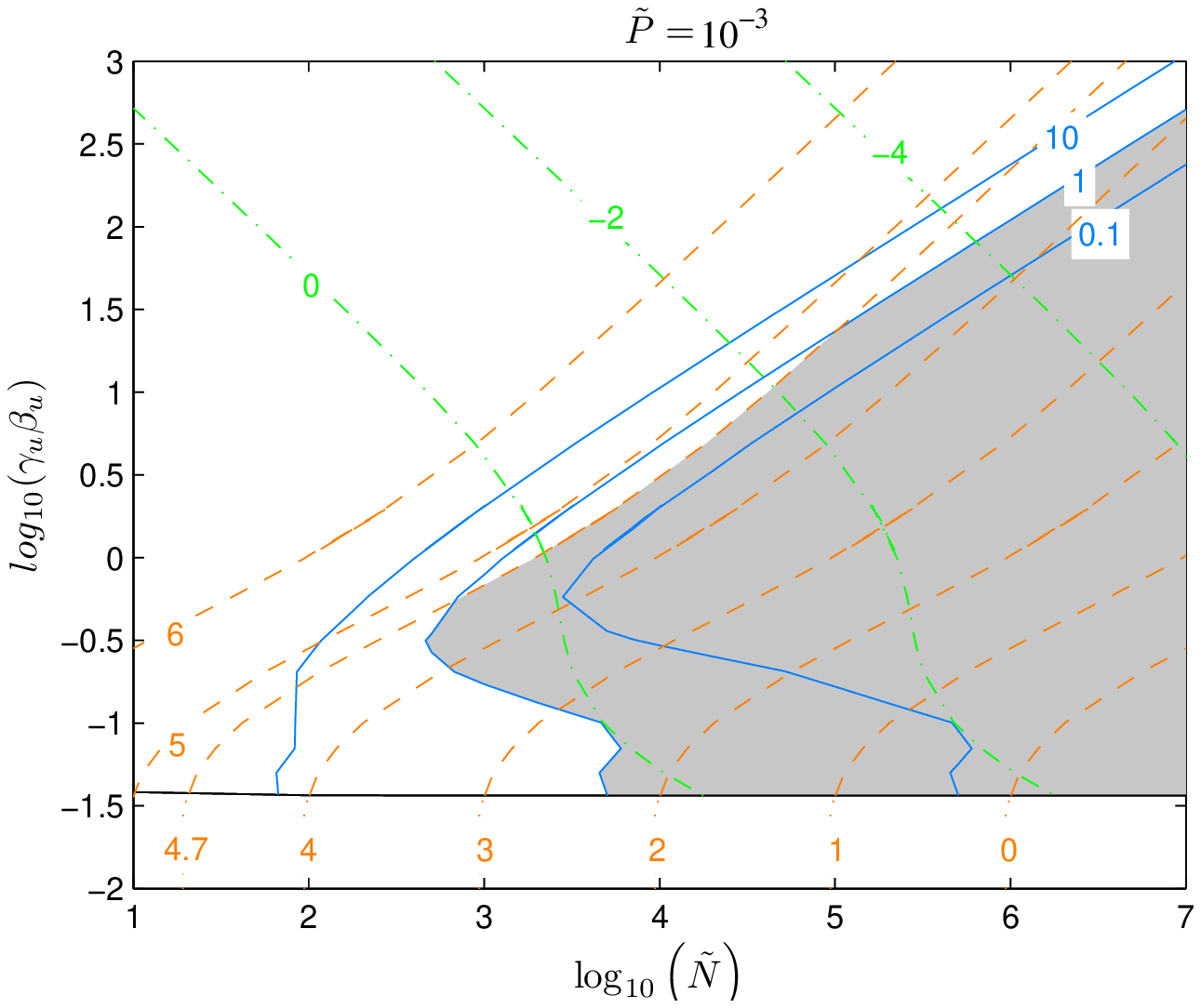} &
\hspace{-10mm}
\includegraphics[width=3in]{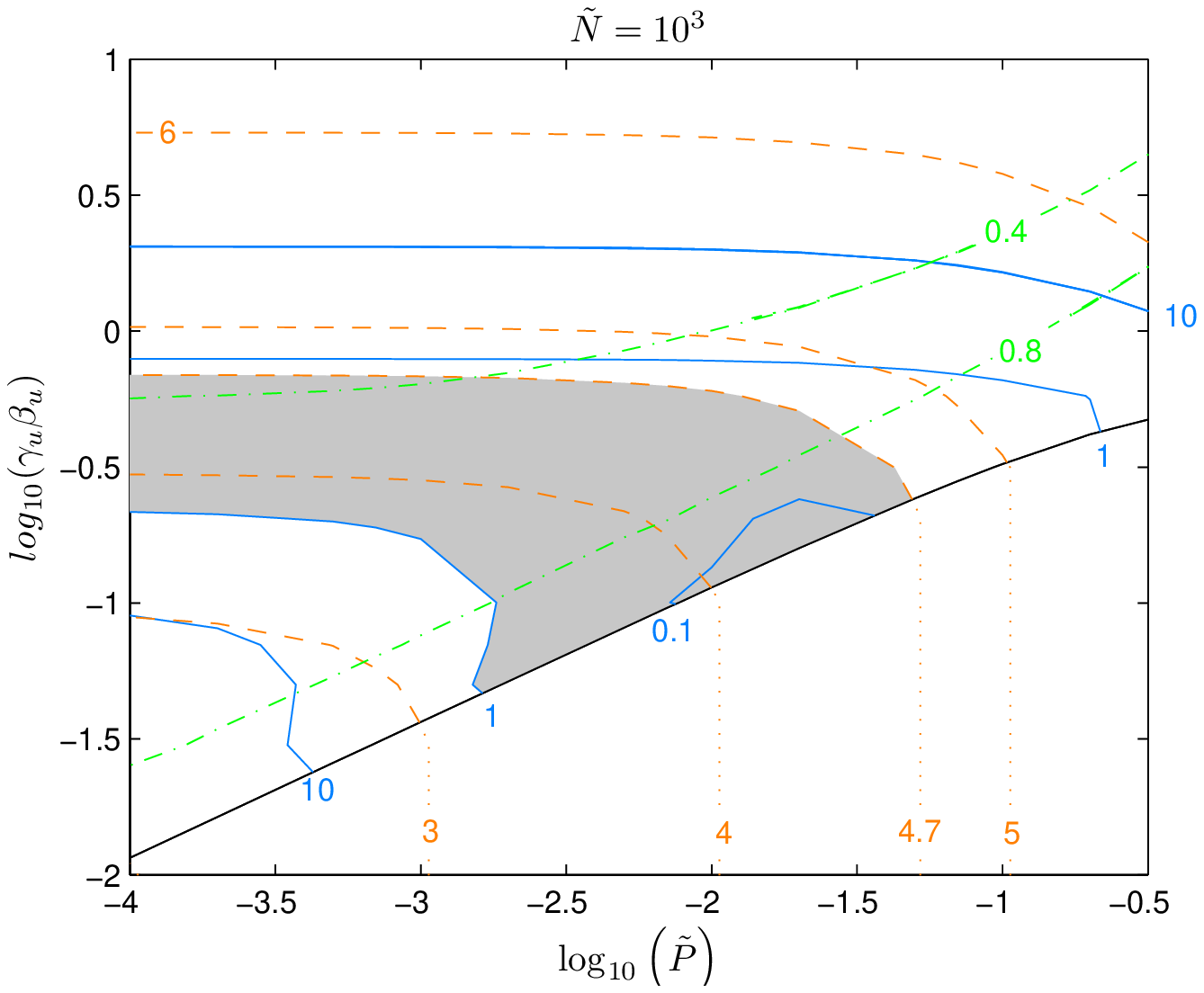} \\
\hspace{-10mm}
\includegraphics[width=3in]{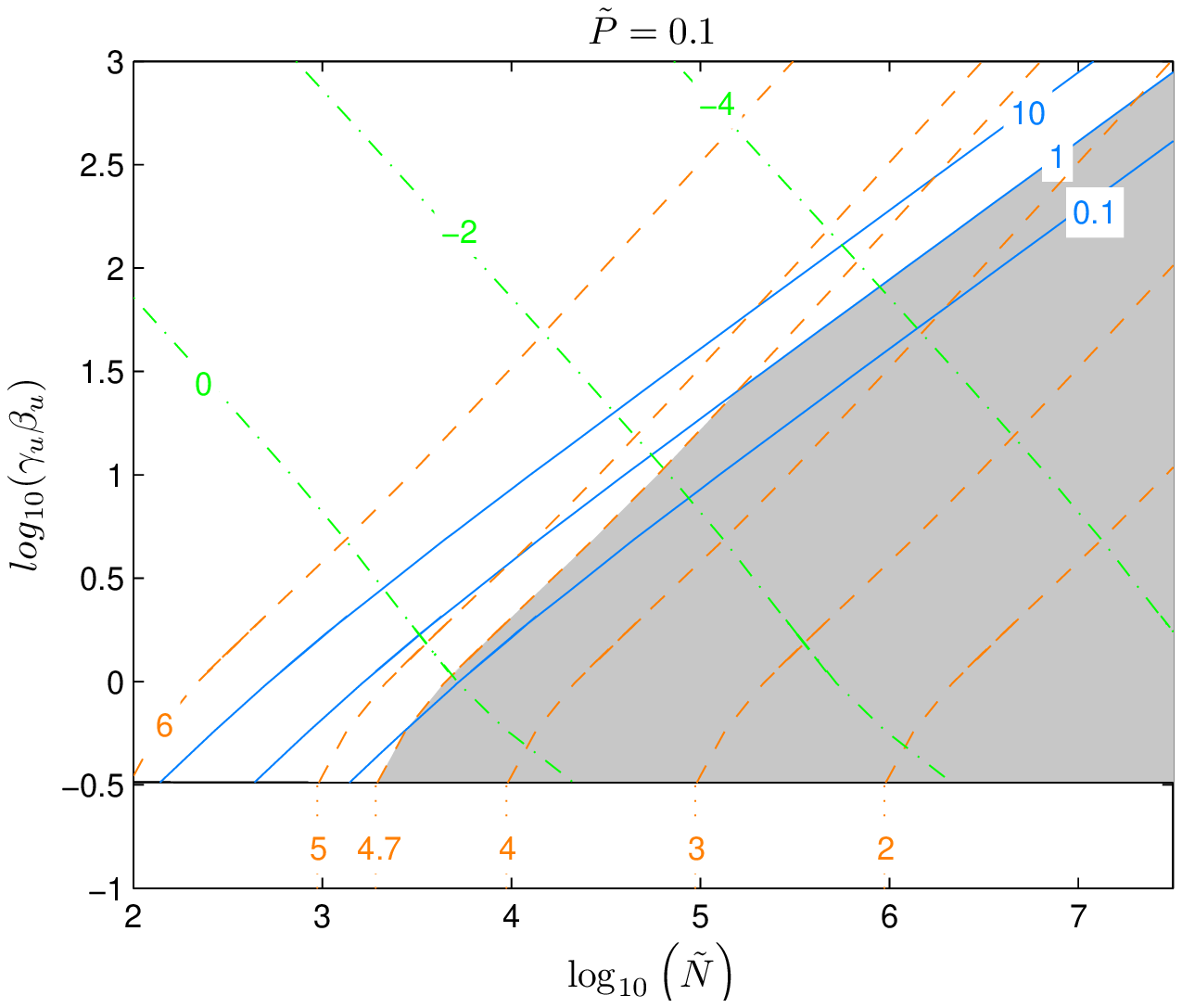} &
\hspace{-10mm}
\includegraphics[width=3in]{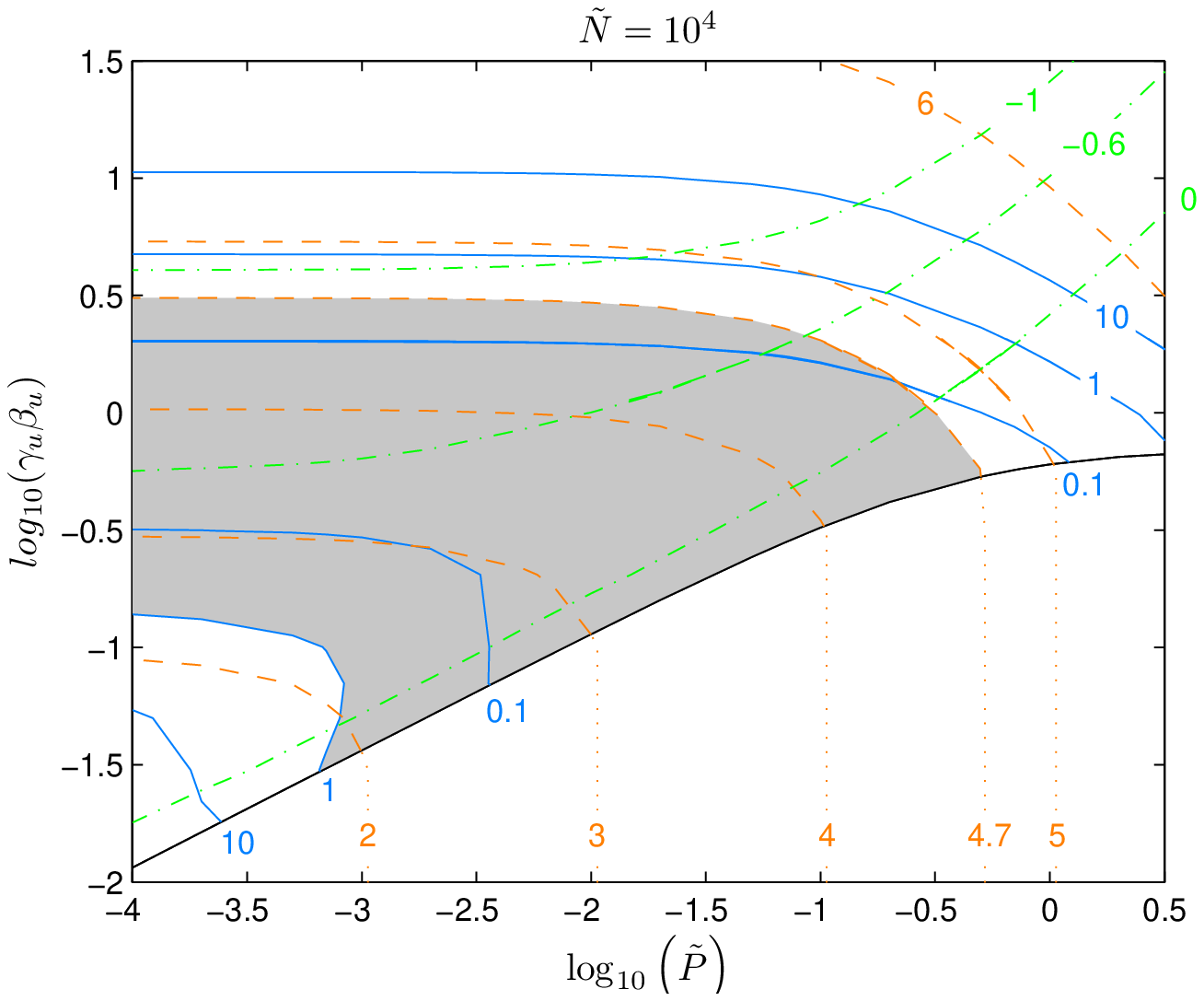} \\
\hspace{-10mm}
\includegraphics[width=3in]{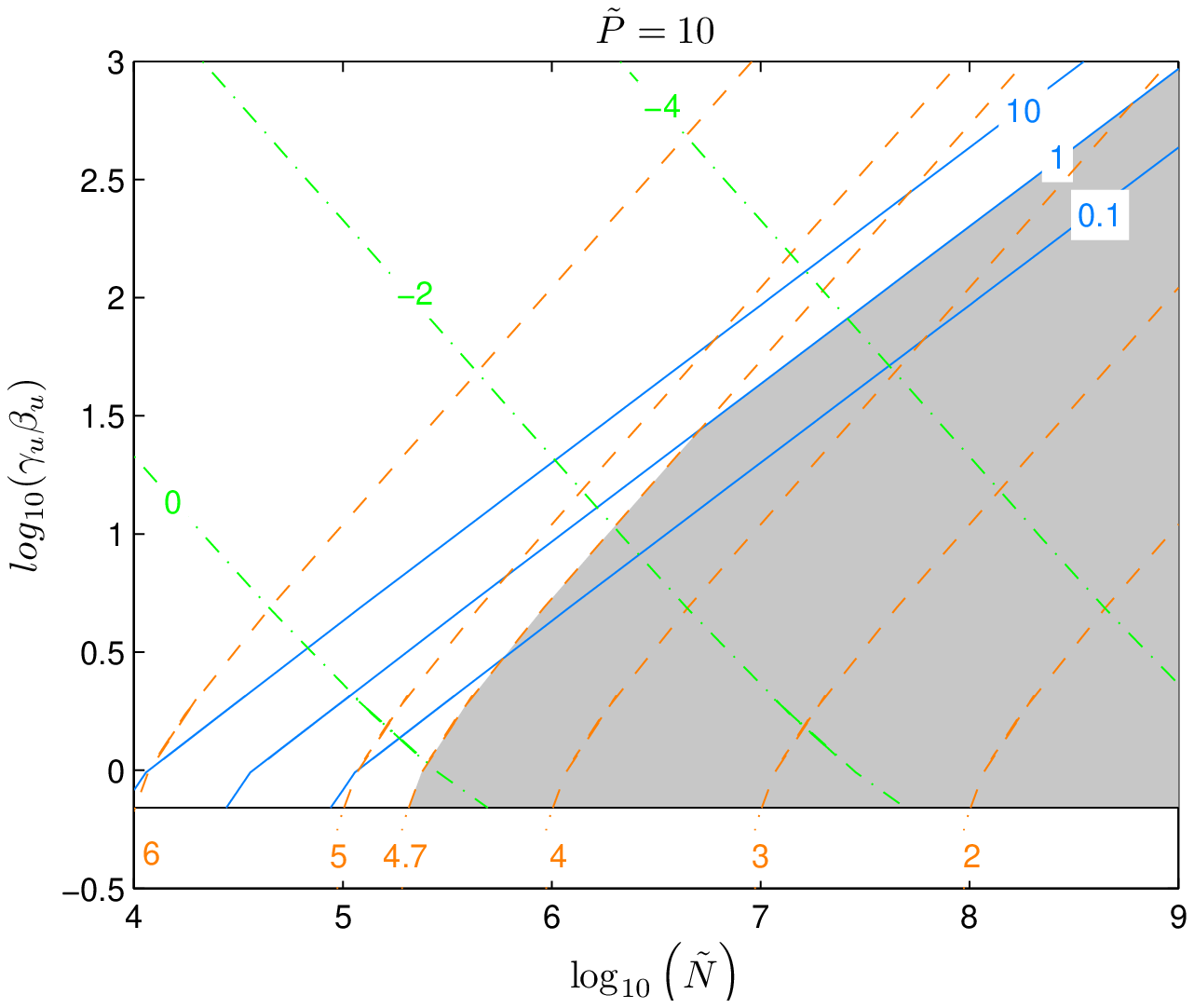} &
\hspace{-10mm}
\includegraphics[width=3in]{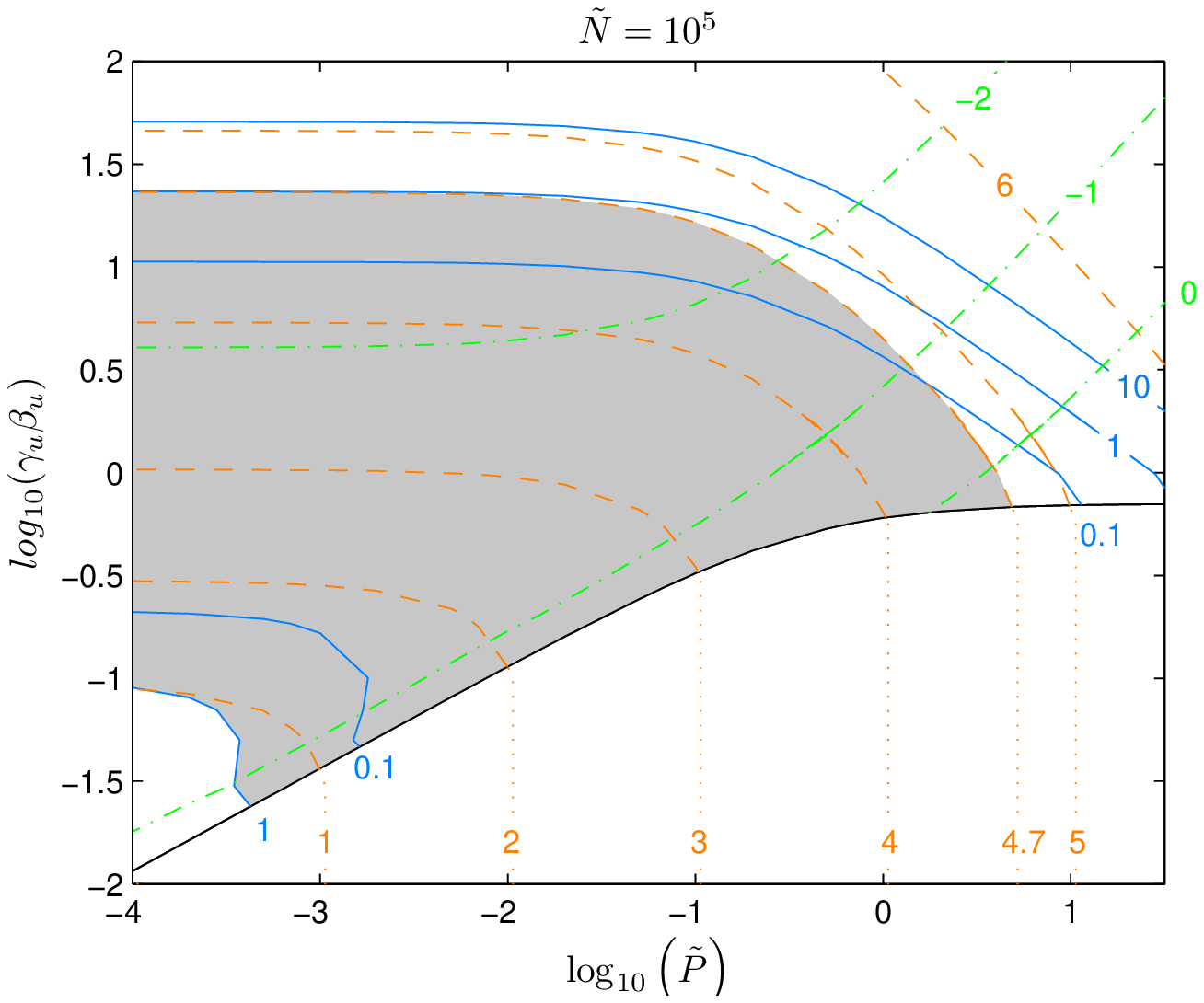}
\end{array}$
\end{center}
\caption{Contour lines of equal $\zeta_c$ (light blue)
and $\log_{10}\left(\frac{KT'_s}{eV}\right)$ (orange), presented
on a $\gamma_u\beta_u$ vs. $\tilde N$ (left panels) and $\tilde P$ (right panels).
Areas marked in gray
correspond those parts in the transition region where the advection
of photons from the upstream dominates over emission processes, and the temperatures are NR. The green line
depicts constant values of $log_{10}(\Upsilon)$, where values $<0$ imply that a high energy
spectral tail may evolve in the shock transition.
Below the black line
shocks can not form since the Mach number $<1$.}\label{zetaNP_fig}
\end{figure}
\thispagestyle{empty}

The numerically calculated values of $T'_s$ and $\zeta_c$ as a
function of $\gamma_u\beta_u$, $\tilde{N}$ and $\tilde{P}$ are
plotted in Fig (\ref{zetaNP_fig}). The figure is divided into 6
panels which represent different cuts in the $\tilde{N}-\tilde{P}$
plane. The left hand panels show plots made with
$\tilde{P}=10^{-3},10^{-1},10$. In the right hand panels we fixed
$\tilde{N}$ to be $\tilde{N}=10^{3},10^4,10^5$ from top to bottom
respectively. The orange lines are contours of equal $T'_s$ in
units of $\log_{10}\left(\frac{KT'_s}{eV}\right)$, whereas the
light blue lines represent contours of equal $\zeta_c$, with
values $\zeta_c=0.1,1,10$. Below the black line the upstream Mach
number $<1$, and no shock occurs.
The region where photon advection from the upstream prevails in the shock transition is
highlighted in gray, and is bounded by $\zeta_c=1$ and
$KT'_s\lesssim0.1m_ec^2$.
At this temperature the energetic photons in the tail of a Wien distributed
radiation constitute $\lesssim1$\% of the total photons number, and their effect
may be ignored.
The real energy distribution may also include a non thermal energetic tail,
where is such a case $KT'_s$ represents the average photon energy rather
than the Wien peak.
This may affect the result, especially in the high $\gamma_u\beta_u$ regime
(see \S\ref{RRMS.spectrum}).
%Even though, since the average energy taken is sufficiently small we expect that it
%will not change the result by much.
The green line describes constant values of $log_{10}(\Upsilon)$, where values $<0$
imply that the photons don't have time to reach thermal equilibrium with the electrons in
the shock transition, and a high energy tail evolves.
This allows for the generation of a high energy tail in the immediate downstream.
It can be seen that there is a considerable
regime of parameters that result in RMS with photon-rich upstreams
which are expected to develop prominent non-thermal tails.

%%%%%%%%%%%%%%%%%%%%%%%%%%%%%%%%%%%%%%%%%%%%%%%%%%%%%%%%%%%%%%%%%%
\subsection {Applications to GRB jets}

Shocks may form in GRB jets due to various reasons, such as
collision between converging flows \citep[e.g.][]{LevEich00,BL07};
interaction of the jet head with a stellar envelope
of a collapsar, prior to its breakout \citep[e.g.][]{Matzner03,Bromberg10};
or collisions of consecutive relativistic shells.
But no matter the source of the dissipation,
if it occurs below the photosphere it will form a RMS.
Here we use the results obtained above to
show that these shocks will most likely have photon-rich
upstreams, and generate photon spectra with
relatively low thermal peaks.
Specifically in the case of shell collisions,
the upstream velocity is mildly relativistic and
the spectrum is expected to have
a hard non-thermal tail.

\begin{figure}
\label{zeta_ph}
\includegraphics{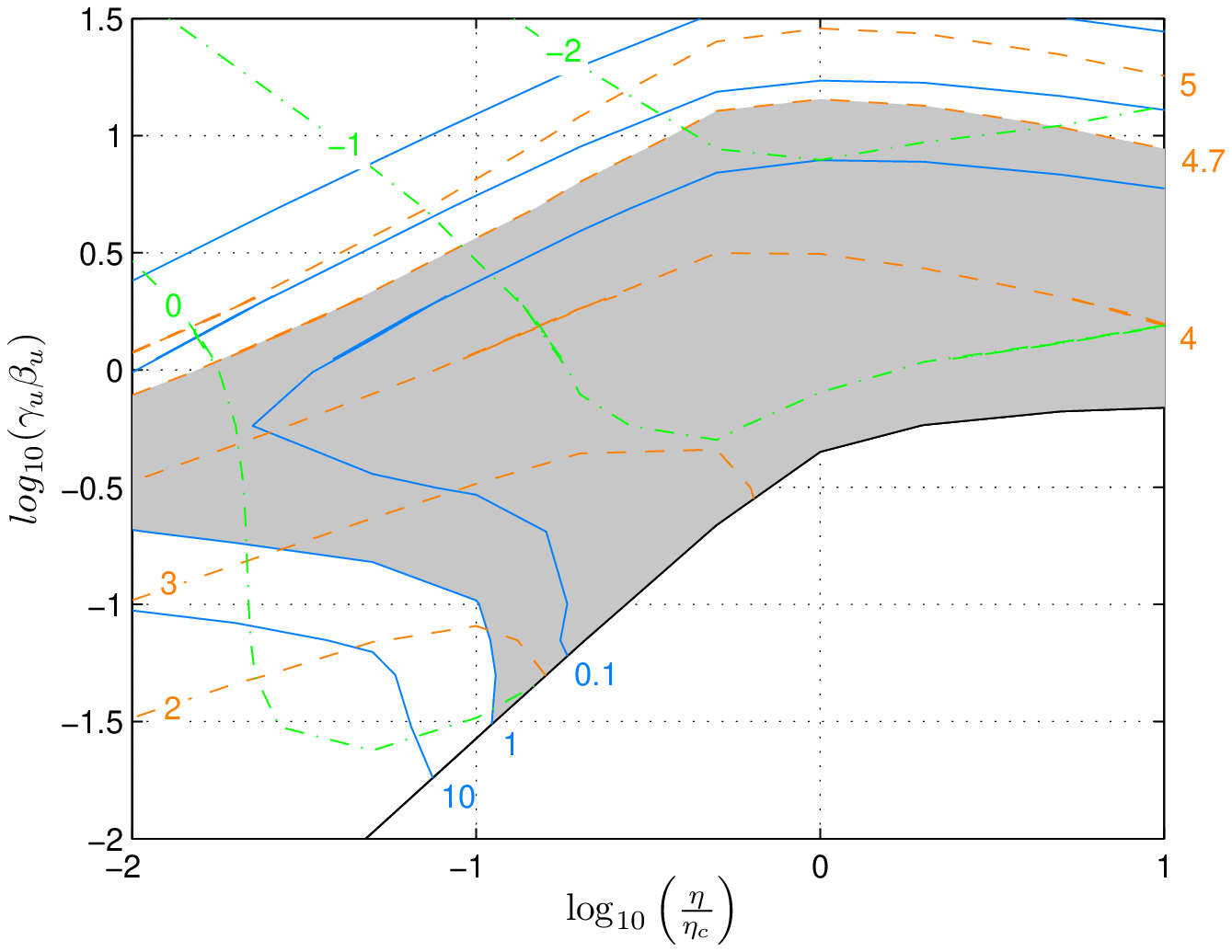}
%\begin{center}
%\hspace{-10mm}
%\epsfig{file=thesis_testn13NR.eps,width=9cm}
%\epsfig{file=thesis_n15NR.eps,width=9cm}
%\hspace{-10mm}
%\epsfig{file=thesis_testn15NR.eps,width=9cm}
%\epsfig{file=thesis_n18NR.eps,width=9cm}
\caption{Contour lines of equal $\zeta_c$ (light blue) and
$\log_{10}\left(\frac{KT'_s}{eV}\right)$ (orange), presented on a
$\gamma_u\beta_u$ vs. $\tildEta/\tildEta_c$ plane. The values here
are calculated on the photosphere. Areas marked in gray correspond
those parts in the transition region where the photons advection dominates
over emission processes, and the temperatures are NR. The green line depicts
constant values of $log_{10}(\Upsilon)$, where values $<0$ imply that a high energy
spectral tail may evolve in the shock transition.
Below the black line shocks can not form since
the Mach number $<1$.}\label{fig_eta}
\end{figure}

Suppose that a shock is formed at some optical depth,
$\tau$, below the photosphere
of a fireball, such as the one discussed in \S3.
The shock upstream condition is determined by
the properties of the fireball at the corresponding dissipation radius.
We can therefore use eq. (\ref{r_ph}) to calculate
the flow parameters, $\tilde{N}$ and $\tilde{P}$ at that
radius giving:
\begin{equation}\label{N_ph}
\tilde{N}_{\tau}=1.25\times10^{5}R_6^{1/4}\frac{\tildEta}{\tildEta_c}\Gamma_0^{
-1/4 },
\end{equation}
\begin{equation}\label{P_ph}
\tilde{P}_{\tau}=\frac{\tau^{1/3}}{4}
\left(\frac{\tildEta}{\tildEta_c}\right)^{4/3}~\left\{
\begin{array}{cc}
\left[3-2\left(\frac{\tildEta_c}{\tildEta}\right)^4\right]^{1/3} &
\tilde\eta_c<\tilde\eta\ll\tildEta_\pm,\\
\tau^{1/3}\left(\frac{\tilde\eta}{\tilde\eta_c} \right )^{4/3} &
\tilde\eta<\tilde{\eta}_c.
\end{array}\right.
\end{equation}
By placing these expressions in eqn. (\ref{Ts_NR})-(\ref{Zeta_R}) we can extract
limits on the values of $\tildEta$ which render the shock upstream photon-rich at that optical depth.
In case where the shock has relativistic upstream, we get:
\begin{equation}
660L_{52}^{1/4}R_6^{-1/2}\gamma_{u,1}\Gamma_0^{1/2}
\lesssim\tildEta\lesssim
1.2\times10^4L_{52}^{1/4}R_6^{1/2}\gamma_{u,1}^{-3}\Gamma_0^{1/2}\tau^{-1}.
\end{equation}
Below and above this regime the temperature at the immediate
downstream exceeds $\sim50~KeV$ which we set as the limit above which
and pairs may affect the flow. In
the non relativistic limit shocks can form only if
$\tildP<0.75\beta_u^2$, corresponding to an upper limit of
\begin{equation}
\tildEta\lesssim 480
L_{52}^{1/4}R_6^{-1/4}\beta_{u,-1}^{3/4}\tau^{-1/4},
\end{equation}
where the lower
limit should be derived numerically. Fig.(\ref{fig_eta}) depicts
the numerically calculated region for $\zeta_c<1$ in the
$\gamma_u\beta_u$ vs. $\tildEta/\tildEta_c$ plane. Here we put
$\tau=1$ and $\Gamma_0=1$ for illustration. The color codes are
similar to fig.(\ref{zetaNP_fig}).

In the standard internal shocks model where shocks from from
shell collisions, $\tau$ and $\eta$ are related via Eq. (\ref{tau}).
The requirement that the shocks form below the photosphere
restricts the jet's baryon load to $\tildEta<\tildEta_{ph}$, defined in
eq. (\ref{eta*}).
This sets limits on $\tildN$ and
$\tildP$, where taking $10^2\eta_2<\tildEta<\tildEta_{ph}$ gives:
\begin{equation}
6.9\cdot10^3L_{52}^{-1/4}R_6^{1/2}\eta_2\Gamma_0^{-1/2}<\tildN_{grb}
<2.5\cdot10^4\left(\frac{b^2-1}{b^2}\frac{R_0}{c\delta t}\right)^{1/5}L_{52}^{-1/20}R_6^{3/10}\Gamma_0^{-1/2},
\end{equation}
and
\begin{equation}
3.1\cdot10^{-3}\left(\frac{b^2-1}{b^2}\frac{R_0}{c\delta t}\right)^{8/15}L_{52}^{-2/15}R_6^{2/15}
<\tildP_{grb}<7.3\cdot10^{-3}\left(\frac{b^2-1}{b^2}\frac{R_0}{c\delta t}\right)^{
2/3}\eta_2^{-2/3},
\end{equation}
where $b=\Gamma_2/\Gamma_1$ is the ratio between the Lorenz factors of the colliding shells,
and $\Gamma_1=\eta$.
If we further assume that the shells have roughly the same mass,
we can parameterize the upstream Lorenz factor at the shock frame.
Equal mass shells have a center of mass which moves with a Lorenz factor
$\Gamma_{CM}=\Gamma_1\sqrt{b}$. Consequently the Lorenz factor of each shell
relative to the central mass is
$\Gamma_{1,CM}=\Gamma_1\Gamma_{CM}\left(1-\beta_1\beta_{CM}\right)=\frac{1+b}{
2\sqrt{b}}$, which for $b>1$ gives an upstream Lorentz factor at the shock frame
\begin{equation}
\gamma_u\lesssim\sqrt{2}\Gamma_{1,CM}=\frac{1+b}{\sqrt{2b}}.
\end{equation}
%Taking $b<6$ for example, gives $\gamma_u\lesssim2$.
Placing these constraints in eq. (\ref{Zeta_R}) gives
\begin{equation}
6\cdot10^{-4}L_{52}^{1/10}R_6^{-3/5}\Gamma_0b^{3/2}\left(\frac{R_0}{c\Delta t}\right)^{2/5}\lesssim
\zeta_c
\lesssim0.01L_{52}^{1/2}R_6^{-1}\eta_2^{-2}\Gamma_0b^{3/2},
\end{equation}
which imply that the shock transition is dominated by the upstream photons,
and its temperature, derived from eq.(\ref{Ts_R}) is:
\begin{equation}
6.5KeV~L_{52}^{1/20}R_6^{-3/10}\sqrt{\Gamma_0b}\left(\frac{c\Delta t}{R_0}\right)^{1/5}\lesssim
KT'_s\lesssim
23KeV~L_{52}^{1/4}R_6^{-1/2}\eta_2^{-1}\sqrt{\Gamma_0b}.
\end{equation}
At such a temperature $\Upsilon\ll1$,
and therefore a hard non-thermal tail may
grow beyond the thermal peak.
Such radiation-mediated internal shocks are therefore expected to exhibit an
observed thermal peak at energies of the order of a few MeV,
together with a prominent hard non-thermal tail which may extend
to a few hundreds MeV, depending on the optical depth at the shock location.
A similar spectrum was observed in the initial phase of GRB 090902B.

%
%
%%%%%%%%%%%%%%%%%%%%%%%%%%%%%%%%%%%%%%%%%%%%%%%%%%%%%%%%%%%%%%%
\section{Bulk Comptonization: Test particle MC simulations}\label{RRMS.spectrum}
%%%%%%%%%%%%%%%%%%%%%%%%%%%%%%%%%%%%%%%%%%%%%%%%%%%%%%%%%%%%%%%
%
%

To illustrate some basic properties of the spectral energy distribution of RRMSs
with photon rich upstreams,
we have preformed a 2D test particle Monte Carlo simulations, using the
shock profiles computed self-consistently by LB08.
For convenience, an analytic fit to these profiles was adopted,
\begin{equation}
U(\xi)= 0.5 \left [\tanh(\alpha \xi)-1 \right ] \left [U_{d} - U_{u} \right ] + U_d.
\label{sime12}
\end{equation}
where $U_u$ is the far upstream 4-velocity, $U_d$ is the far downstream 4-velocity and
$\alpha$ is a free parameter that determines the width of the transition layer.   The
dimensionless variable  $\xi$ is related to the angle averaged optical depth across the shock. In terms of the
coordinate $x$ it is expressed as

\begin{equation}
\xi = n_{u}\sigma_{T} x,
\label{sime13}
\end{equation}
where $n_u$ is proper density far upstream and $\sigma_T$ is the Thomson cross section.
As stated above, these global shock solutions ignores KN effects and pair productions, and are suitable
in the regime where convection of seed photons by the upstream flow is sufficiently large.   As will be shown below,
the neglect of pair production may not be justified at all for large enough upstream Lorentz factors.
We have kept the full KN cross section in our test particle MC simulations to examine its effect on the
spectrum, but ignored the thermal spread of the electrons in the shock and downstream.   Thus, the results
are relevant for photons above the thermal peak.

The injected photons are drawn from a monochromatic and isotropic (in the fluid rest frame) source of photons,
located at a fixed position within the simulation box, as viewed in the shock frame.
We have run simulations for different source locations, including injection from the
immediate upstream and immediate downstream, and found little differences in the resultant
spectrum as long as the optical depth from the injection point to the shock transition is
not too large. The boundaries of the simulation box are defined by two values of the variable
$\xi$, $\xi_{min}~\mbox{and}~\xi_{max}$, where $\xi_{min}$ is located upstream and $\xi_{max}$
downstream.  The boundaries $\xi_{min}$ and $\xi_{max}$
are chosen to optimize the run time without affecting the resultant spectrum. That is, it has
to be large enough to allow injected photons to undergo the maximum number of scattering across
the shock, but minimum number of scattering in the far downstream, where the probability to
recross the shock becomes very small.  The location of $\xi_{min}$ is chosen such that less then 2\%
of the injected photons escape the simulation box though the upstream boundary.

\begin{figure}
%\label{shock_spec1}
%\includegraphics{downP7P1vfvb.eps}
\includegraphics{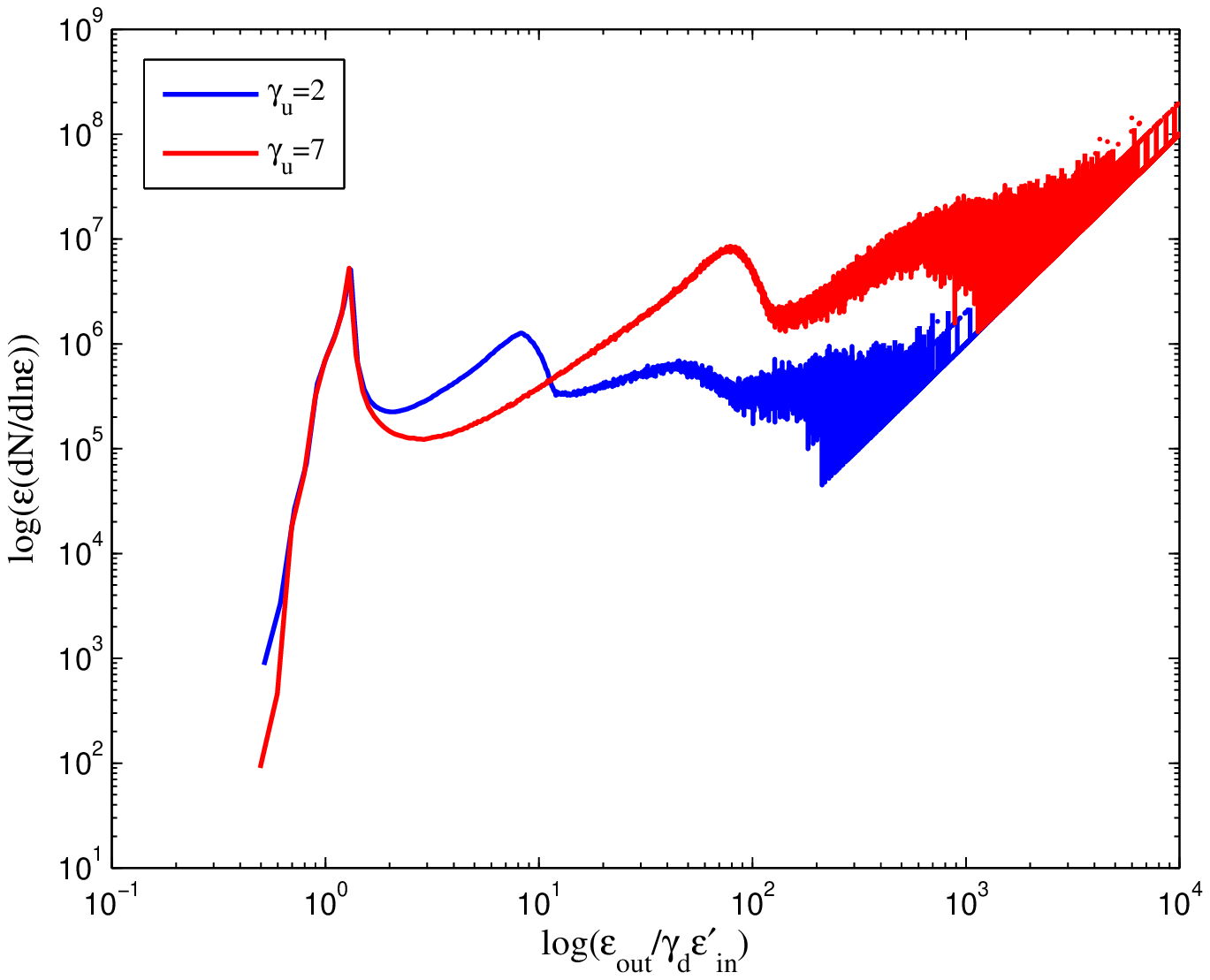}
\caption{\label{shock_spec1}The angle averaged spectral energy distribution
in the immediate downstream of a RRMS of width $\tau=3$ and Lorentz factor
$\gamma_u=2$ (red) and $\gamma_u=7$ (blue), as measured in the shock frame.
The downstream velocity in both cases is taken to be $\beta_d=1/3$.
The calculated energy is normalized to the initial injection energy, $\gamma_d\epsilon'_{in}$,
with $\epsilon'_{in}=100$ ev in the fluid frame. }
\end{figure}

\begin{figure}
%\label{shock_spec2}
%\includegraphics{downG2difftaub.eps}
\includegraphics{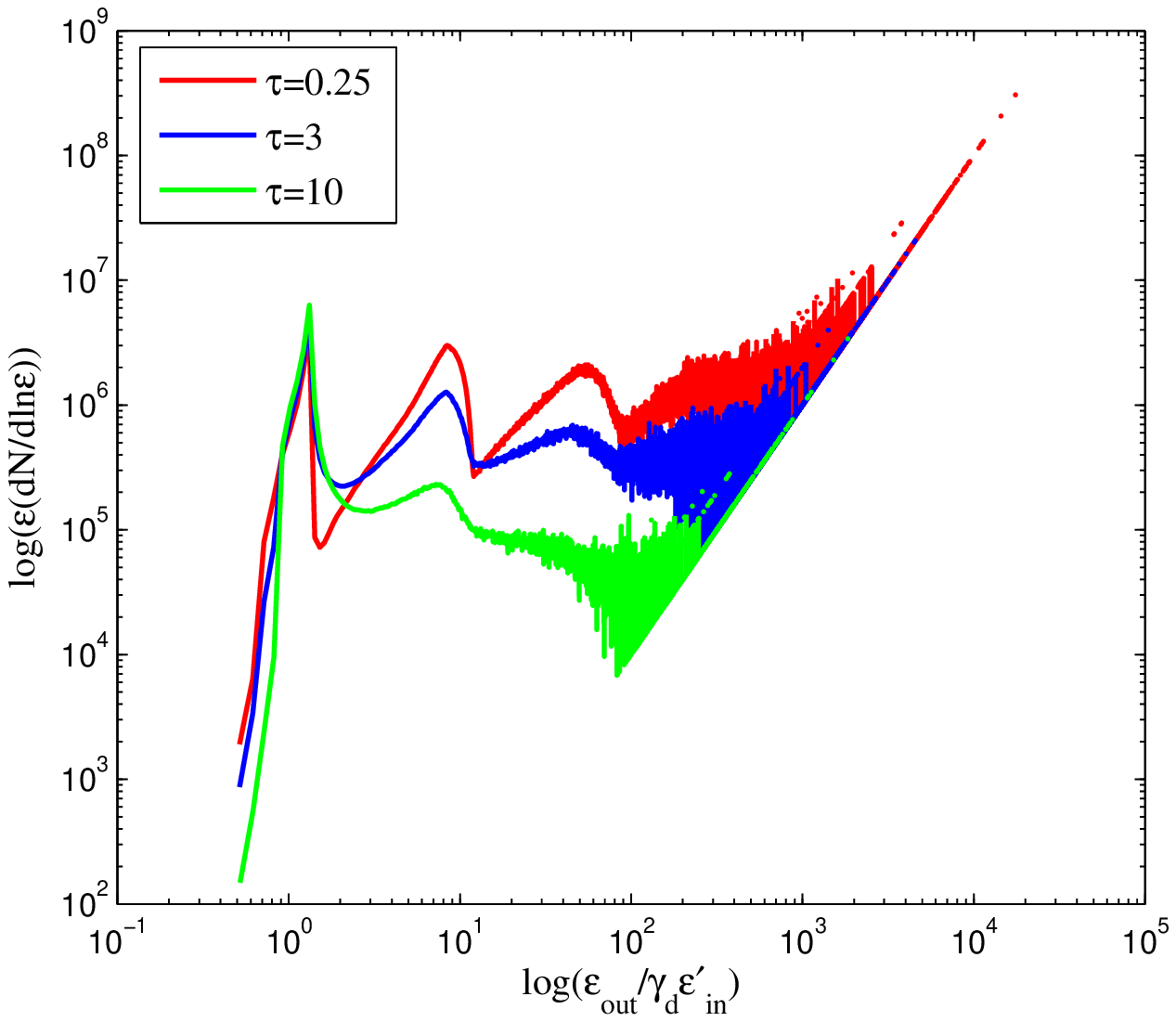}
\caption{\label{shock_spec2}Same as Fig \ref{shock_spec1}, but for an upstream Lorentz factor $\gamma_u=2$ and
different values of $\tau$.}
\end{figure}

For each of the results presented in Figs. (\ref{shock_spec1}) and (\ref{shock_spec2}) below $10^7$ photons were injected
in the immediate downstream.  The trajectory of each photon
was followed until it ``escaped'' the simulation box from either side.
Fig. (\ref{shock_spec1}) delineates the effect of
the upstream Lorentz factor on the spectrum at the immediate downstream.  The width of the
shocks in both cases is $\tau=3$, which roughly gives the profile computed in LB08,
where $\tau$ is the Thomson depth for a photon moving from the downstream, at
$U(\xi_d)=U_d(1+10^{-4})$, to the point $U(\xi_u)=0.9U_u$ upstream, viz.,
\begin{equation}
\tau=\int_{\xi_u}^{\xi_d}\gamma(\xi)(1+\beta)d\xi,
\end{equation}
with $\gamma(\xi)=[U^2(\xi)+1]^{1/2}$ and $\beta=U/\gamma$ for the profile given in Eq. \ref{sime12}.
As seen, the spectrum hardens with increasing Lorentz factor, but even for modest ones we find that a considerable
fraction of the shock energy is transferred to photons near the KN limit.
The effect of the shock width on the photon spectrum is explored in Fig. \ref{shock_spec2}.
We have also performed runs for different injection points, including photon injection upstream, and
found only small differences in the resulted spectra.  We conclude that the location where photons
are injected does not affect the spectrum considerably.

In all cases studied a hard spectrum was obtained, extending over several decades in energy. For sufficiently high
injection energies the statistics allowed us to follow the spectrum up to the KN limit, $\gamma_um_ec^2$, as
measured in the shock frame.  There is no evidence for a cutoff or steepening at energies well below the KN limit,
in any of the experiments performed.  As naively expected, the spectrum is somewhat harder for larger upstream Lorentz factors.
There is also a weak dependence on the shock width.   The profiles computed in LB08 are relatively narrow,
in the sense that the angle averaged optical depth of the order of a few ($\tau\sim3-5$).  As demonstrated by \citet{Budnik10}, pair production
upstream can lead to a photon breading cycle that can boost the upper cutoff well above the KN limit, up to an energy
of $\gamma_u^2m_ec^2$ in case of  sufficiently relativistic shocks.
The temperature in the immediate downstream in
the case of photon rich upstream is much smaller than in the case of a cold upstream studied by \citet{Budnik10}.
Nevertheless our test particle simulations
illustrate that pair production by the accelerated photons is expected to be important, especially in the case of moderate to high upstream Lorenz
factor, and might alter the details of the resultant spectrum.

%
%
%%%%%%%%%%%%%%%%%%%%%%%%%%%%%%%%%%%%%%%%%%%%%%%%%%%%%%%%%%%%%%%%%
\section{Conclusions}
%%%%%%%%%%%%%%%%%%%%%%%%%%%%%%%%%%%%%%%%%%%%%%%%%%%%%%%%%%%%%%%%%

In this paper we consider sub-photospheric emission by relativistic radiation mediated shocks
in the prompt phase of GRBs.   The main conclusions are:

1. Dissipation via internal shocks is likely to occur over a range of radii that encompasses the photosphere.
Slower shells should collide below the photosphere, at a moderate optical depth (unless the time separation between consecutive ejections
is much larger than the dynamical time of the engine), whereas shells ejected with sufficiently large Lorentz factor
will collide above the photosphere.    The fraction of the energy that dissipates below the photosphere depends on
the burst parameters, and may significantly vary from burst to burst.    Specific examples, for which measurements  of the
luminosity and estimates of the Lorentz factors are available were analyzed.   In case of
GRB 090902B we conclude that a major fraction of the explosion energy dissipated below the photosphere, in
a region of optical depths $<300$, via relativistic radiation mediated shocks.   For GRB 080916C we find that
collision of shells occurred predominantly above the photosphere, giving rise to formation of collisionless shocks.

2. Shocks that form below the photosphere are mediated by Compton scattering of radiation produced inside the
shock or advected by the upstream flow.   Since the scale of the shock, a few Thomson lengths, is vastly larger
than any kinetic scale involved, particle acceleration by the Fermi process is highly unlikely in such shocks.
On the other hand, bulk Comptonization gives rise to a hard spectrum extending above a (thermal) Wien peak, up to the
KN limit $\gamma_um_ec^2$ (and perhaps beyond, e.g., Budnik et al. 2010),
where $\gamma_u$ is the Lorentz factor of the upstream flow, measured in the shock frame. The mean photon energy in the
immediate downstream is determined by the number of photons advected into and/or produced inside the shock.
For a cold upstream, a condition expected in shock breakout episodes, the Wien temperature is $\sim m_ec^2$ (Budnik et al. 2010).
However, under the conditions anticipated in the prompt phase of GRBs,
the shock upstream is photon rich, in the sense that  photon advection largely dominates over photon production,
and the thermal peak is located at energies well below $m_ec^2$.  We estimate temperatures in the range between
a few and a few tens keV for typical GRB parameters.

The existence of the hard spectral component is demonstrated in this work
through test particle Monte Carlo simulations on given shock profiles, that were computed
in an earlier work (Levinson and Bromberg 2008), neglecting pair production by MeV photons and the thermal spread of the
electrons. The resultant spectra extend to the KN limit, which imply that pair production inside the shock
is likely to be important in photon rich shocks.  A full treatment requires a proper account of this process
in the computation of the shock structure and the resultant spectrum.

\subsection*{Acknowledgments}
We thank Ehud Nakar and Re'em Sari for enlightening discussions.
This research was supported by an ERC advanced research
grant, and by by an ISF grant for the
Israeli center for high energy astrophysics.

\newpage
\pagestyle{empty}

\end{document}